 \documentclass[doublespacing]{elsart}

 \usepackage{graphics}
 \usepackage{graphicx}

\usepackage{amssymb}

\begin{document}

\begin{frontmatter}



\title{DC SQUID based on the mesoscopic multiterminal Josephson junction}

 \author[a]{M.H.S. Amin}
 \author[b]{A.N. Omelyanchouk \corauthref{cor1}}
 \author[a,c]{A.M.Zagoskin}
 \address[a]{D-Wave Systems Inc., 320-1985 W. Broadway,Vancouver, B.C., V6J 4Y3, Canada}

 \address[b]{B.I.Verkin Institute for Low Temperature
Physics and Engineering,Ukrainian National Academy of Sciences,
Lenin Ave. 47, Kharkov 310164, Ukraine }
\address[c]{Physics and Astronomy Dept., The University of
British Columbia,6224 Agricultural Rd., Vancouver, B.C., V6T 1Z1,
Canada}

\corauth[cor1]{Corresponding author, omelyanchouk.ilt.kharkov.ua}

\begin{abstract}
A theory is offered for a novel device, mesoscopic four-terminal
SQUID. The studied system consists of  a  mesoscopic four-terminal
junction, one pair of terminals of which is incorporated in a
superconducting ring and the other one is connected with a
transport circuit. The nonlocal weak coupling between the
terminals leads to effects of phase dragging and magnetic flux
transfer. The behaviour of  a four-terminal SQUID, controlled by
the external parameters, the applied magnetic flux and the
transport current is investigated. The critical current and the
current voltage characteristics as functions of magnetic flux are
calculated. In the nonlocal mesoscopic case they depend not only
on the magnitude of the applied flux but also on its sign,
allowing measurement of the direction of the external magnetic
field.
\end{abstract}

\end{frontmatter}

The four-terminal SQUID is a superconducting device based on a
four-terminal Josephson junction \cite{oov95}. The multiterminal
Josephson junctions (MTJ) generalizes the usual (two-terminal)
Josephson junctions \cite{bar} to the case of weak coupling
between several massive superconducting banks (terminals).
Compared with two-terminal junctions, such systems have additional
degrees of freedom and the corresponding set of control
parameters, preset transport currents and (or) applied magnetic
fluxes. As a result, the current- or voltage-biased and the
magnetic flux-driven regimes can be combined in one multiterminal
microstructure. One of the realizations of a MTJ presents  the set
of superconducting terminals, which are connected with the short
dirty microbridges having a common centre (Fig.1a). In such
(conventional) MTJ the nonlinear coupling between individual
bridges is due to the fact that the superconducting order
parameter in the common centre is a function of the currents
through all the microbridges. See review of macroscopic quantum
interference effects in conventional multiterminal microstructures
in Ref.\cite{oo}. Another type of multiterminal Josephson junction
is based on the weak coupling of bulk superconductors through the
two-dimensional clean normal layer (Fig.1b). In such a mesoscopic
ballistic 4-terminal junction \cite{zo} the nonlocal coupling of
supercurrents is established due to the phase dependent local
Andreev levels inside the weak link. The nonlocal weak coupling in
mesoscopic MTJ leads to the effects of phase dragging and magnetic
flux transfer \cite{oz,aoz1}. In the present paper we consider the
behaviour of a mesoscopic 4-terminal SQUID (Fig.2) emphasizing
the specific features of mesoscopic case.

The studied system (Fig.2) consists of a mesoscopic four-terminal
junction, one pair of terminals of which is incorporated in a
superconducting ring and the other one is connected with a
transport circuit. The system can be controlled by an external
magnetic flux $\Phi^e$ and by the transport current $I$. Current
$J$ is the self-induced circulating current in the ring circuit
with self-inductance $L$. The state of the 4-terminal junction is
determined by the dynamical variables, phases $\varphi_{i}$ of the
complex off-diagonal potential $\Delta_{0}\exp({\rm i}\varphi_i)$
in the \textit{i}-th terminal (\textit{i}=1...4). In the
4-terminal case we have three independent phase differences.
Phases $\theta$ and $\phi$ are the phase differences between
terminals 2-1 and 3-4, respectively. Phase $\chi=1/2(\varphi
_{1}+\varphi _{2})-1/2(\varphi _{3}+\varphi _{4})$ is the phase
difference between the ring and the transport current circuit. The
phase $\phi$ is related to the observable quantity, total magnetic
flux threading the ring $\Phi$, $\phi=\frac{2e}{\hbar}\Phi $.  For
given external parameters $I$ and $\Phi^e$, Gibbs potential for
the 4-terminal SQUID in terms of variables $\phi,\theta$ and
$\chi$ has the form \cite{aoz1}:
\begin{eqnarray}
  && U(\phi,\theta,\chi ;I,\Phi^e) = \kappa \frac{(\phi-\frac{2e}{\hbar}\Phi^e)^2}
  {2\widetilde{L}} - I \theta -\cos(\theta)- \kappa \cos(\phi) \nonumber \\
  &&- 2 \left({\alpha} \cos{\theta\over 2}\cos {\phi \over
  2}-{\beta}
  \sin {\theta\over 2} \sin {\phi\over 2} \right) \cos(\chi).
\end{eqnarray}
The last three terms in Eq.(1) are the Josephson coupling energy.
The coefficients $\kappa,\alpha, \beta$ in Eq.(1) determine the
coupling between the different terminals. They depend on the
geometry of the weak link and on the transparency $D$ of $S-N$
interfaces. In a symmetric case of square $N$-layer and ideal
transparency ($D$=1) they equal to : $\kappa=1,\alpha=\sqrt{2}+1,
\beta=\sqrt{2}-1$. The current $I$ is measured in units of
$I_0=\pi\gamma_{12}{\Delta_0(T)}^2/4eT_c$, where $\gamma_{12}$ is
the inverse Sharvin resistance between the terminals 1 and 2.
$\widetilde{\L} = (2e/\hbar)LI_0\kappa$ is the dimensionless
self-inductance.

The specific feature of a mesoscopic 4-terminal SQUID is the
appearance in the potential $U$ of the term, proportional to
coefficient $\beta$. Note, that even for a completely symmetrical
mesoscopic MTJ we have $\beta\neq 0$, contrary to the conventional
MTJ (Fig.1a). In the latter case $\beta$ identically equals zero
for any symmetry of microbridge configuration.

The minimization of $U$ with respect to phases $\theta,\phi,\chi$
gives the equations
\begin{equation}
I=\sin{\theta}+ \left[\alpha
\sin{\frac{\theta}{2}}\cos{\frac{\phi}{2}}+ \beta
\cos{\frac{\theta}{2}}\sin{\frac{\phi}{2}} \right]\cos{\chi},
\end{equation}
\begin{equation}
\frac {\frac{2e}{\hbar}\Phi^e-\phi} {{\widetilde{L}
}}=\sin{\phi}+\frac{1}{\kappa} \left[\alpha
\sin{\frac{\phi}{2}}\cos{\frac{\theta}{2}}+ \beta
\cos{\frac{\phi}{2}}\sin{\frac{\theta}{2}} \right]\cos{\chi},
\end{equation}
\begin{equation}
\cos{\chi}={\rm sign} \left[\alpha \cos{\frac{\phi}{2}}
\cos{\frac{\theta}{2}}-\beta \sin{\frac{\phi}{2}}
\sin{\frac{\theta}{2}} \right],
\end{equation}
The system of Eqs.(2-4) describes the behaviour of a mesoscopic
4-terminal SQUID in the stationary state. It determines the
magnetic flux in the ring $\Phi=\frac{\hbar}{2e}\phi$ as function
of applied magnetic flux $\Phi^e$ and transport current $I$.The
nonlocal Josephson coupling ($\beta\neq 0$) leads to the new
effects. As follows from Eqs.(2-4) the magnetic flux $\Phi$ in the
ring produces the phase difference $\theta$ on current driven
junction 1-2 at zero  current $I$ flowing from terminal 1 to
terminal 2. Similarly the transport current $I$ induces in the
ring the magnetic flux $\Phi$ in the absence of external flux
$\Phi^e$. The influence of the phase difference between one pair
of terminals on the phase difference between another pair of
terminals is what we call phase dragging effect in the mesoscopic
4-terminal junction. As applied to mesoscopic 4-terminal SQUID
this effect results in the dependence of the critical current
$I_c$ (maximal value of $I$ at which the system of equations (2-4)
has the solution) not only on the absolute value  but also on the
sign of the applied magnetic flux $\Phi^e$. The steady states
domain in the $(I,\Phi^e)$ plane is shown in Fig.3 for the case of
small self-inductance  $\widetilde{L} \ll 1 $. The boundary of the
domain , curve $I_c(\Phi^e)$ (solid lines in Fig.3), is $2\pi$
periodic, but due to the terms proportional to $ \beta $ in
Eqs.(2-4), it is not invariant under the transformation $\Phi^e
\rightarrow -\Phi^e$. The symmetry is restored if we
simultaneously change $\Phi^e$ on $-\Phi^e$ and $I$ on $-I$. Note,
that in conventional case ($\beta=0$) the boundary of the steady
state domain $I_c(\Phi^e)$ is symmetric with respect to the axes
$(I,\Phi^e)$ (dashed line in Fig.3). Outside the steady state
domain, the stationary solutions for $(\theta,\Phi)$ are absent
and system goes to the nonstationary resistive regime. We have
studied the dynamical behaviour of the 4-terminal SQUID in the
frame of the heavily damped resistively shunted junction (RSJ)
model \cite{bar}. The features of the dynamical behaviour of the
mesoscopic 4-terminal SQUID are again affected by the terms
proportional to $\beta$, i.e. by nonlocal coupling. The
current-voltage characteristics in the transport channel, $V(I)$,
($V$ is the time averaged voltage between terminals 2 and 1) are
shown  in Fig.4 for different values of $\Phi^e$.  The voltage
$V(I)$ in applied magnetic flux $\Phi^e$ depends on the sign of
the $\Phi^e$ as well as critical current $I_c$. Thus, the
mesoscopic 4-terminal SQUID can be used for direction-sensitive
detection of weak magnetic fluxes. In addition, the system
described by the potential (1) permits the existence of the
bistable states even for negligible dimensionless self-inductance
 $\widetilde{L}$ $ \rightarrow 0 $, which can be exploited in the
superconducting phase qubit design (see Ref.\cite{qub}).

\newpage

\section*{Figure Captions}

\begin{itemize}

\item[Fig.1:]{The multiterminal Josephson junction. Bulk
superconductors 1,..4 are connected with the crossed microbridges
(a) or are weakly coupled through the rectangular of a
two-dimensional electron gas (b).}
\item[Fig.2:]{The mesoscopic four-terminal SQUID.}
\item[Fig.3:]{The steady state domain for mesoscopic four-terminal
SQUID in plane $(I,\Phi_e)$ of the control parameters (solid
line). $\Phi^e$ is measured in units $\hbar/2e$. Dashed line
corresponds to the conventional four-terminal SQUID.}
\item[Fig.4:]{The current-voltage characteristics of a mesoscopic 4-terminal SQUID
for different values of applied flux $\Phi^e$.
$\Phi_0=\frac{h}{2e}$ is the superconducting flux quantum,
$\widetilde{L}=0$.}
\end{itemize}
\newpage
\begin{figure}[!t]
\begin{center}
\includegraphics[width=0.9 \textwidth]{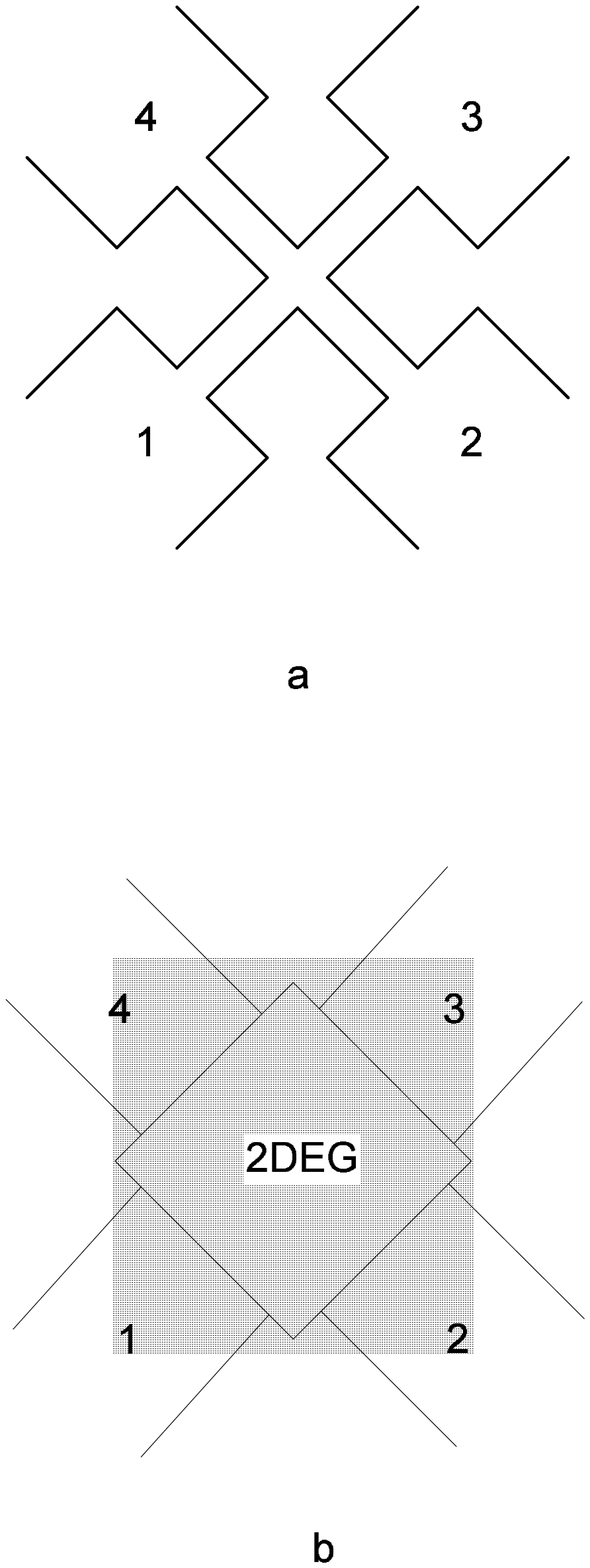}
\end{center}
\caption{{}}
\end{figure}
\newpage
\begin{figure}[!t]
\begin{center}
\includegraphics[width=0.9 \textwidth]{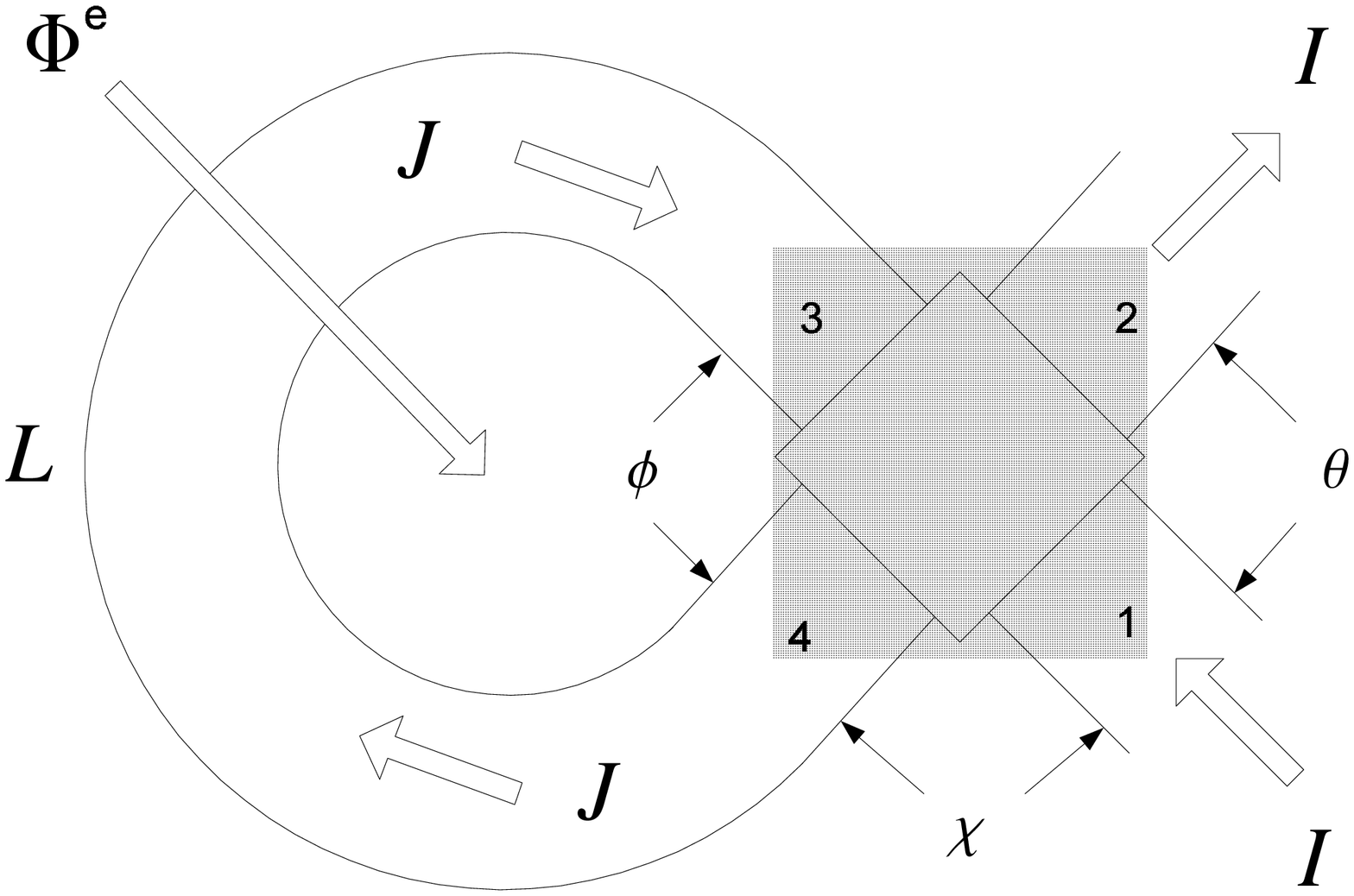}
\end{center}
\caption{{}}
\end{figure}
\newpage
\begin{figure}[!t]
\begin{center}
\includegraphics[width=1 \textwidth]{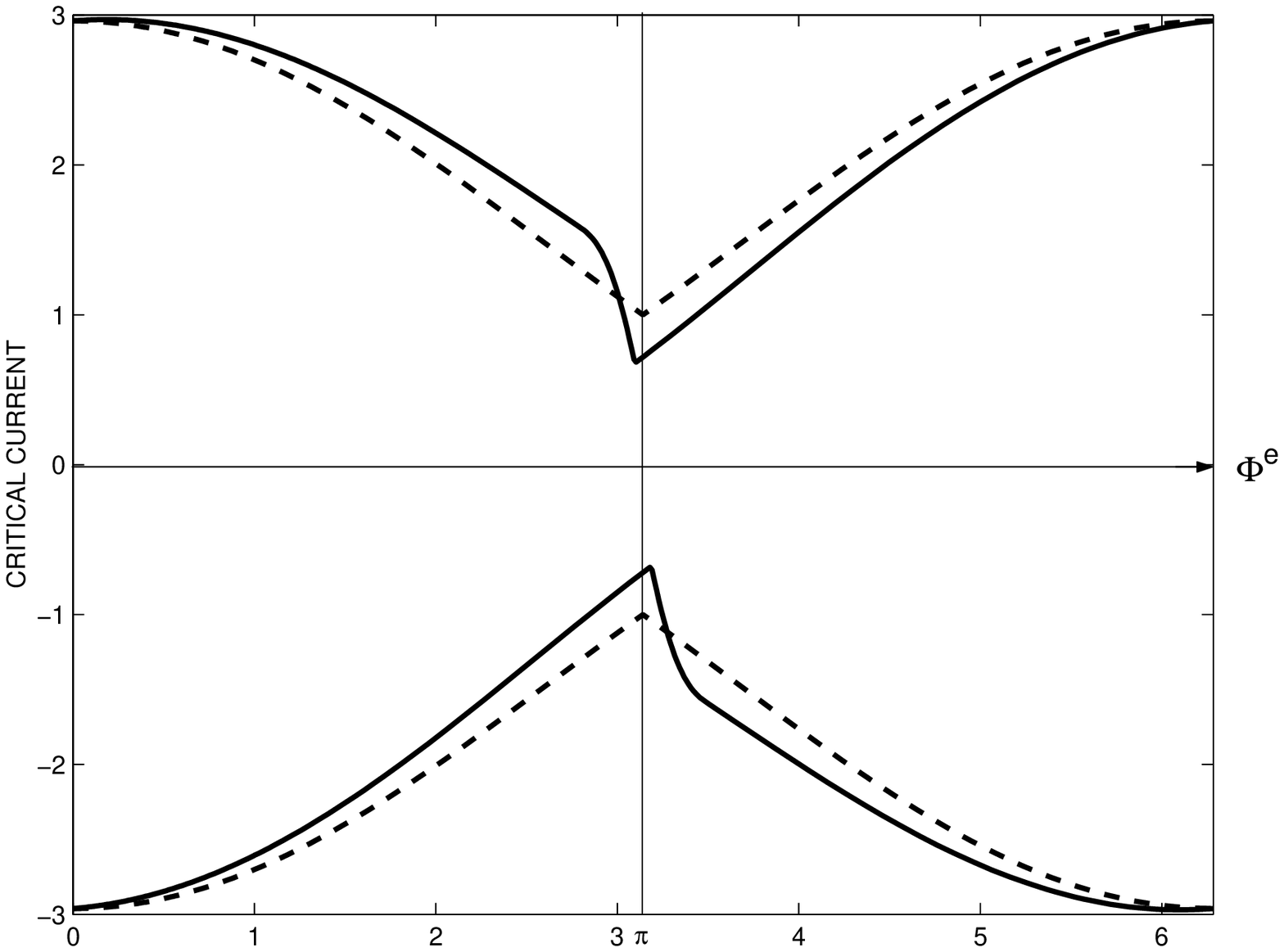}
\end{center}
\caption{{}}
\end{figure}
\newpage
\begin{figure}[!t]
\begin{center}
\includegraphics[width=1 \textwidth]{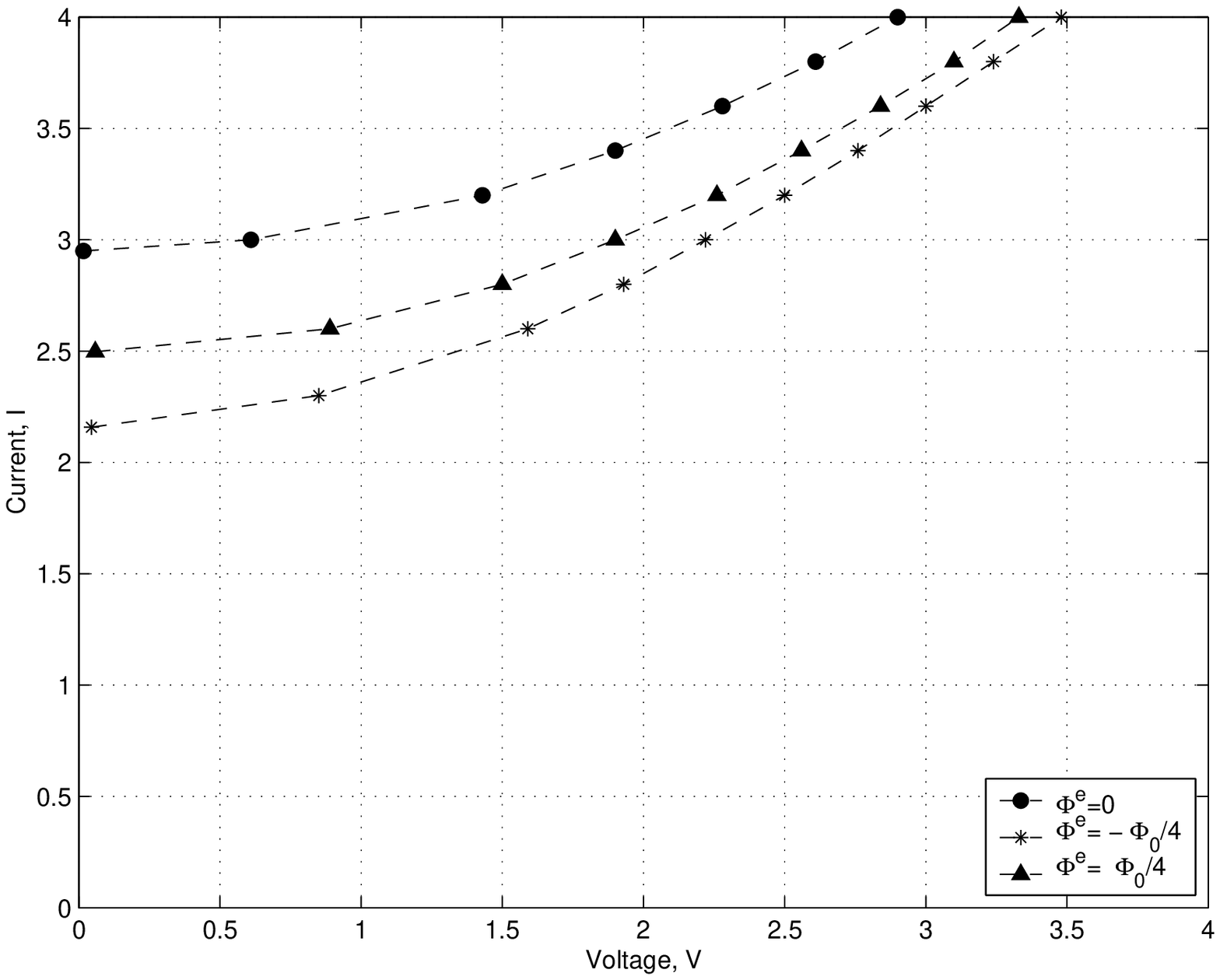}
\end{center}
\caption{{}}
\end{figure}

\begin{thebibliography}{99}
\bibitem{oov95}  R. de Bruyn Ouboter, A.N.Omelyanchouk, E.D.Vol,
Physica B 205 (1995) 153.
\bibitem{bar}  A.Barone, G.Paterno, Physics and Applications of the
Josephson Effect, Wiley, New York, 1982.
\bibitem{oo} R. de Bruyn Ouboter, A.N.Omelyanchouk, Superlatt. {\&}
Microstruct., {\bf 23}(1999)1005.
\bibitem{zo} Malek Zareyan and A.N.Omelyanchouk,Low Temp.Phys. \textbf{25}
(1999) 175.
\bibitem{oz}A.N.Omelyanchouk and Malek Zareyan, Physica B 291 (2000)
81.
\bibitem{aoz1} M.H.S.Amin, A.N.Omelyanchouk and A.M.Zagoskin, Low
Temp. Phys. \textbf{27}, N8, (2001) 616.
\bibitem{qub} M.H.S.Amin, A.N.Omelyanchouk, A.Blais, A.Maassen van den Brink, G.Rose, T.Duty
and A.M.Zagoskin, Multi-terminal superconducting phase qubit
(these proceedings).
\end{thebibliography}
\end{document}